\documentclass[12pt, twoside]{article}

\usepackage{a4wide, amsmath, latexsym, epsfig,
  psfrag,graphicx, rotating, fancyhdr, tabularx, afterpage}

\usepackage{feynarts}
\usepackage{axodraw}

\newcommand{\id}{{\rm 1\kern-.12em
\rule{0.3pt}{1.5ex}\raisebox{0.0ex}{\rule{0.1em}{0.3pt}}}}

\newcommand{\seff}{\sin^2\theta_{\rm eff}}

\begin{document}

\thispagestyle{empty}
\setcounter{page}{0}
\def\thefootnote{\fnsymbol{footnote}}

{\textwidth 15cm

\begin{flushright}
MPP-2005-109\\
hep-ph/0509302 \\
\end{flushright}

\vspace{2cm}

\begin{center}

{\Large\sc {\bf Higgs-mass dependence of the
                effective \\[0.1cm]
           electroweak mixing angle  \boldmath{$\sin^2\theta_{\rm eff}$}
\\[0.3cm]
  at the two-loop level
               }}

\vspace{2cm}

{\sc W. Hollik, U. Meier} and 
{\sc S. Uccirati~\footnote{Address after October 3 2005, 
INFN Sezione di Torino, Italy}}

\vspace*{1cm}

     Max-Planck-Institut f\"ur Physik \\[0.1cm] 
     (Werner-Heisenberg-Institut)\\[0.1cm]
     D-80805 M\"unchen, Germany
\end{center}

\vspace*{2cm}

\begin{abstract}
\noindent 
The result for the Higgs-dependent 
electroweak two-loop bosonic contributions to 
the effective leptonic mixing angle of the $Z$-boson 
in the Standard Model is presented. 
Together with the previously calculated fermionic contributions
it yields the complete dependence of $\sin^2\theta_{\rm eff}$
on the Higgs-boson mass $M_H$. 
Compared to the fermionic contributions,
the bosonic contributions are found to be smaller and have   
the opposite sign, compensating part of the fermionic contributions.

\end{abstract}

}
\def\thefootnote{\arabic{footnote}}
\setcounter{footnote}{0}

\newpage

\section{Introduction}

The effective electroweak mixing angle for leptons, $\seff$,
is experimentally determined with high accuracy
from measurements of various asymmetries on the $Z$ resonance, 
with the current value  $0.23153 \pm 0.00016$~\cite{unknown:2004qh}, 
and  further improvements are expected from future collider 
experiments~\cite{Aguilar-Saavedra:2001rg,Baur:2001yp}.  
Analyses done in combination with the theoretical predictions
for $\seff$ in the Standard Model
yield stringent bounds on the Higgs-boson mass $M_H$, making $\seff$
a precision observable of central importance for tests of the 
Standard Model. Therefore, a precise
and reliable calculation is a necessity, requiring at least
the complete electroweak two-loop contributions.

$\seff$ is determined from the ratio of the dressed vector and axial vector couplings 
$g_{V,A}$ of the $Z$ boson to leptons~\cite{Bardin:1997xq},
\begin{eqnarray}
\seff &=& \frac{1}{4}  \left( 1-\mathrm{Re}\, \frac{g_V}{g_A} \right) . \label{s2w}
\end{eqnarray}
It is related to the vectorboson-mass ratio or the on-shell quantity $s_W^2$, respectively,
via
\begin{eqnarray}
\seff &=& \kappa \, s_W^2 \; = \; 
\kappa \,\left( 1-\frac{M_W^2}{M_Z^2} \right) \, , \quad  
 \kappa = 1+\Delta\kappa \, , \label{kappa}
\end{eqnarray}
involving the $\kappa$ factor, which is unity at the tree level and accommodates the 
higher-order contributions in $\Delta\kappa$.
In recent independent calculations
the complete two-loop electroweak corrections of the fermionic type, i.e.\
with at least one closed fermion loop, to $\Delta\kappa$ were 
obtained~\cite{Awramik:2004ge, Hollik:2005va}. 
The bosonic two-loop corrections, however, are still missing.
In this note we present a first step 
towards the full $\mathcal{O}\left(\alpha^2\right)$ 
result for $\Delta\kappa$, namely the results of the subclass of Higgs dependent
contributions, thus providing the complete 
Higgs-boson mass dependence of the bosonic two-loop corrections.

\section{Structure of electroweak two-loop contributions}\label{2loop}

\begin{figure}[!b]
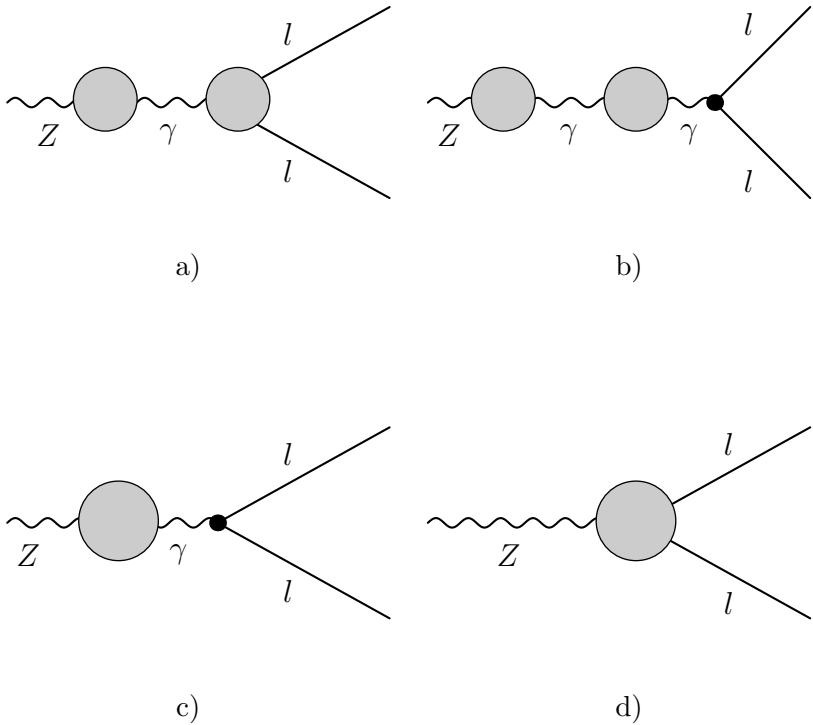

 
\begin{feynartspicture}(350,350)(2,2.2)

\FADiagram{a)}
\FAProp(0.,10.)(11.,10.)(0.,){/Sine}{0}
\FALabel(9.,7.43)[t]{$\gamma$}
\FALabel(3.,7.43)[t]{$Z$}
\FAProp(20.,15.)(11.,10.)(0.,){/Straight}{0}
\FALabel(15.2273,11.3749)[br]{$l$}
\FAProp(20.,5.)(11.,10.)(0.,){/Straight}{0}
\FALabel(15.2273,5.62506)[tr]{$l$}
\FAVert(11.,10.){0}
\GCirc(110.,239.){12.}{0.8}
\GCirc(60.,239.){12.}{0.8}

\FADiagram{b)}
\FAProp(0.,10.)(15.,10.)(0.,){/Sine}{0}
\FALabel(13.,7.43)[t]{$\gamma$}
\FALabel(7.,7.43)[t]{$\gamma$}
\FALabel(1.,7.43)[t]{$Z$}
\FAProp(20.,15.)(15.,10.)(0.,){/Straight}{0}
\FALabel(16.2273,11.8749)[br]{$l$}
\FAProp(20.,5.)(15.,10.)(0.,){/Straight}{0}
\FALabel(16.2273,5.12506)[tr]{$l$}
\FAVert(15.,10.){0}
\GCirc(260.,239.){12.}{0.8}
\GCirc(210.,239.){12.}{0.8}

\FADiagram{c)}
\FAProp(0.,10.)(11.,10.)(0.,){/Sine}{0}
\FALabel(2.,8.43)[t]{$Z$}
\FALabel(9.5,8.43)[t]{$\gamma$}
\FAProp(20.,15.)(11.,10.)(0.,){/Straight}{0}
\FALabel(15.2273,12.3749)[br]{$l$}
\FAProp(20.,5.)(11.,10.)(0.,){/Straight}{0}
\FALabel(15.2273,6.62506)[tr]{$l$}
\FAVert(11.,10.){0}
\GCirc(65.,80.){15.}{0.8}

\FADiagram{d)}
\FAProp(0.,10.)(11.,10.)(0.,){/Sine}{0}
\FALabel(4.,8.43)[t]{$Z$}
\FAProp(20.,15.)(11.,10.)(0.,){/Straight}{0}
\FALabel(15.2273,12.8749)[br]{$l$}
\FAProp(20.,5.)(11.,10.)(0.,){/Straight}{0}
\FALabel(15.2273,6.02506)[tr]{$l$}
\FAVert(11.,10.){0}
\GCirc(260.0,80.){15.}{0.8}

\end{feynartspicture}

\hspace*{-10cm}
\caption{Generic classes of two loop diagrams}
\label{gendiag}
\end{figure}

The general strategy of our calculation of the two-loop electroweak 
contributions to $\seff$, including renormalization, has been described
in~\cite{Hollik:2005va}. As outlined in~\cite{Hollik:2005va}, 
one has to take into account basically the classes of diagrams 
depicted schematically in Fig.~\ref{gendiag}, where the circles denote 
renormalized two- and three-point irreducible vertex functions at the
one-loop level in Fig.~\ref{gendiag}a and at two-loop order in
Fig.~\ref{gendiag}c and \ref{gendiag}d.

The real part of the diagram shown in Fig.~\ref{gendiag}c vanishes in the on-shell 
renormalization scheme~\cite{Freitas:2002ja} adopted 
in our calculation and 
the diagrams of Fig.~\ref{gendiag}a and \ref{gendiag}b
only contribute products of imaginary parts of one-loop functions. 
So we are left with the irreducible two-loop
$Z\ell\ell$-vertex diagrams in Fig.~\ref{gendiag}d.
The $Z$ couplings in (\ref{s2w}) appear in the renormalized $Z\ell\ell$ vertex
for on-shell $Z$ bosons,
\begin{eqnarray}
\hat{\Gamma}_\mu^{Z\ell\ell\,(2)}(M^2_Z) &=& \gamma_\mu\,
   \left( g_V - g_A \gamma_5 \right) \, . 
\end{eqnarray}
As for the fermionic corrections, we split the two-loop contribution 
for the renormalized vertex into two UV-finite pieces according to 
\begin{eqnarray}
\hat{\Gamma}_\mu^{Z\ell\ell\,(2)}(M^2_Z) &=& 
\Gamma_\mu^{Z\ell\ell\,(2)} (M^2_Z) + \Gamma_\mu^{CT}
\nonumber \\[0.2cm]
 &=&  
\left[ \Gamma_\mu^{Z\ell\ell\, (2)}(0)+ \Gamma_\mu^{CT}\right]
+\left[ \Gamma_\mu^{Z\ell\ell\,(2)}(M^2_Z) - 
\Gamma_\mu^{Z\ell\ell\,(2)}(0) \right] .\label{split}
\end{eqnarray}
$\Gamma_\mu^{Z\ell\ell\,(2)}\left(P^2\right)$ 
denotes the corresponding unrenormalized  $Z\ell\ell$ vertex
for on-shell leptons and momentum transfer $P^2$,
and $\Gamma_\mu^{CT}$ is the two-loop counter term, which is independent of $P^2$.
The first term in the decomposition of~(\ref{split}) therefore 
contains the complete two-loop renormalization, 
but no genuine two-loop vertex diagrams since in absence of external 
momenta they reduce to simpler vacuum integrals. 
All the genuine two-loop vertex diagrams appear as subtracted quantities in
the second term in~(\ref{split}).

\begin{figure}[!htb]
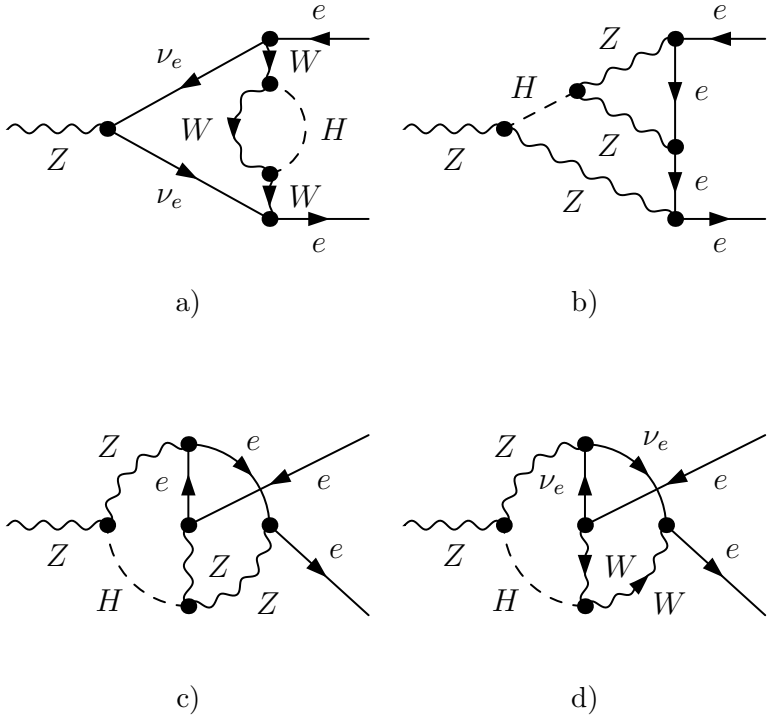

\begin{feynartspicture}(300,350)(2,2.3)

\FADiagram{a)}
\FAProp(0.,10.)(5.5,10.)(0.,){/Sine}{0}
\FALabel(2.75,8.93)[t]{$Z$}
\FAProp(20.,15.)(14.5,15.)(0.,){/Straight}{1}
\FALabel(17.25,16.07)[b]{$e$}
\FAProp(20.,5.)(14.5,5.)(0.,){/Straight}{-1}
\FALabel(17.25,3.93)[t]{$e$}
\FAProp(5.5,10.)(14.5,15.)(0.,){/Straight}{-1}
\FALabel(9.72725,13.3749)[br]{$\nu_e$}
\FAProp(5.5,10.)(14.5,5.)(0.,){/Straight}{1}
\FALabel(9.72725,6.62506)[tr]{$\nu_e$}
\FAProp(14.5,12.5)(14.5,15.)(0.,){/Sine}{-1}
\FALabel(15.57,13.75)[l]{$W$}
\FAProp(14.5,7.5)(14.5,12.5)(0.8,){/ScalarDash}{0}
\FALabel(17.32,10.)[l]{$H$}
\FAProp(14.5,7.5)(14.5,12.5)(-0.8,){/Sine}{-1}
\FALabel(11.43,10.)[r]{$W$}
\FAProp(14.5,7.5)(14.5,5.)(0.,){/Sine}{1}
\FALabel(15.57,6.25)[l]{$W$}
\FAVert(14.5,7.5){0}
\FAVert(14.5,12.5){0}
\FAVert(5.5,10.){0}
\FAVert(14.5,15.){0}
\FAVert(14.5,5.){0}

\FADiagram{b)}
\FAProp(0.,10.)(5.5,10.)(0.,){/Sine}{0}
\FALabel(2.75,8.93)[t]{$Z$}
\FAProp(20.,15.)(15.,15.)(0.,){/Straight}{1}
\FALabel(17.5,16.07)[b]{$e$}
\FAProp(20.,5.)(15.,5.)(0.,){/Straight}{-1}
\FALabel(17.5,3.93)[t]{$e$}
\FAProp(5.5,10.)(15.,5.)(0.,){/Sine}{0}
\FALabel(10.0081,6.60838)[tr]{$Z$}
\FAProp(9.55,12.1)(5.5,10.)(0.,){/ScalarDash}{0}
\FALabel(7.40659,11.7241)[br]{$H$}
\FAProp(9.55,12.1)(15.,15.)(0.,){/Sine}{0}
\FALabel(12.0269,14.4383)[br]{$Z$}
\FAProp(15.,9.)(9.55,12.1)(0.,){/Sine}{0}
\FALabel(11.9886,9.6827)[tr]{$Z$}
\FAProp(15.,9.)(15.,15.)(0.,){/Straight}{-1}
\FALabel(16.07,12.)[l]{$e$}
\FAProp(15.,9.)(15.,5.)(0.,){/Straight}{1}
\FALabel(16.07,7.)[l]{$e$}
\FAVert(15.,9.){0}
\FAVert(9.55,12.1){0}
\FAVert(5.5,10.){0}
\FAVert(15.,15.){0}
\FAVert(15.,5.){0}

\FADiagram{c)}
\FAProp(0.,10.)(5.5,10.)(0.,){/Sine}{0}
\FALabel(2.75,8.93)[t]{$Z$}
\FAProp(20.,15.)(10.,10.)(0.,){/Straight}{1}
\FALabel(17.1318,12.7564)[tl]{$e$}
\FAProp(20.,5.)(14.5,10.)(0.,){/Straight}{-1}
\FALabel(17.8126,8.16691)[bl]{$e$}
\FAProp(10.,5.5)(5.5,10.)(-0.4,){/ScalarDash}{0}
\FALabel(6.41076,6.41076)[tr]{$H$}
\FAProp(10.,5.5)(10.,10.)(0.,){/Sine}{0}
\FALabel(11.07,7.75)[l]{$Z$}
\FAProp(10.,5.5)(14.5,10.)(0.4,){/Sine}{0}
\FALabel(13.766,6.23398)[tl]{$Z$}
\FAProp(10.,14.5)(5.5,10.)(0.4,){/Sine}{0}
\FALabel(6.23398,13.766)[br]{$Z$}
\FAProp(10.,14.5)(10.,10.)(0.,){/Straight}{-1}
\FALabel(8.93,12.25)[r]{$e$}
\FAProp(10.,14.5)(14.5,10.)(-0.4,){/Straight}{1}
\FALabel(13.1968,14.1968)[bl]{$e$}
\FAVert(10.,14.5){0}
\FAVert(10.,5.5){0}
\FAVert(5.5,10.){0}
\FAVert(10.,10.){0}
\FAVert(14.5,10.){0}

\FADiagram{d)}
\FAProp(0.,10.)(5.5,10.)(0.,){/Sine}{0}
\FALabel(2.75,8.93)[t]{$Z$}
\FAProp(20.,15.)(10.,10.)(0.,){/Straight}{1}
\FALabel(17.1318,12.7564)[tl]{$e$}
\FAProp(20.,5.)(14.5,10.)(0.,){/Straight}{-1}
\FALabel(17.8126,8.16691)[bl]{$e$}
\FAProp(10.,5.5)(5.5,10.)(-0.4,){/ScalarDash}{0}
\FALabel(6.41076,6.41076)[tr]{$H$}
\FAProp(10.,5.5)(10.,10.)(0.,){/Sine}{-1}
\FALabel(11.07,7.75)[l]{$W$}
\FAProp(10.,5.5)(14.5,10.)(0.4,){/Sine}{1}
\FALabel(13.766,6.23398)[tl]{$W$}
\FAProp(10.,14.5)(5.5,10.)(0.4,){/Sine}{0}
\FALabel(6.23398,13.766)[br]{$Z$}
\FAProp(10.,14.5)(10.,10.)(0.,){/Straight}{-1}
\FALabel(8.93,12.25)[r]{$\nu_e$}
\FAProp(10.,14.5)(14.5,10.)(-0.4,){/Straight}{1}
\FALabel(13.1968,14.1968)[bl]{$\nu_e$}
\FAVert(10.,14.5){0}
\FAVert(10.,5.5){0}
\FAVert(5.5,10.){0}
\FAVert(10.,10.){0}
\FAVert(14.5,10.){0}

\end{feynartspicture}
\caption{Examples of diagrams containing internal Higgs bosons}
\label{Higgsdiag}
\end{figure}

As a first step towards the complete bosonic two-loop corrections, 
we consider the Higgs-mass dependence. To this end we calculate
the subtracted two-loop bosonic $\Delta\kappa^{(\alpha^2)}_{bos}$,
\begin{eqnarray}
\Delta\kappa^{(\alpha^2)}_{bos,sub}  &=& 
\Delta\kappa^{(\alpha^2)}_{bos}\left(M_H\right)-\Delta\kappa^{(\alpha^2)}_{bos}\left(M_H^0\right), 
\label{subtract}
\end{eqnarray}
for a fixed reference mass of the Higgs boson, 
chosen as $M^0_H = 100$ GeV. 
The quantity $\Delta\kappa^{(\alpha^2)}_{bos,sub}$
is UV finite and gauge-parameter independent. 
The dependence on $M_H$ enters exclusively through diagrams with internal 
Higgs boson lines. 
Some typical examples are displayed in Fig.~\ref{Higgsdiag}.

The computation of the renormalized vertex at $P^2=0$ [first term in~(\ref{split})] 
can be done in analogy to the fermionic case~\cite{Hollik:2005va}, which means
generating Feynman diagrams with {\it FeynArts}~\cite{Hahn:2000jm} and applying 
{\it TwoCalc}~\cite{Weiglein:1993hd} to reduce the amplitudes to 
standard integrals. The resulting scalar one-loop integrals and two-loop vacuum integrals
are calculated using  known
analytic results~\cite{'tHooft:1978xw,Davydychev:1992mt}. The two-loop self-energies 
with non-vanishing external momentum, as part of the two-loop counterterm, 
are obtained with the help of
one-dimensional integral representations~\cite{Bauberger:1994nk}. 
Moreover, we implemented new methods described in~\cite{Ferroglia:2003yj},
and used them for cross checks.

For the subtracted vertex, the 
second term in equation~(\ref{split}), two independent calculations were performed, 
based either on
{\it FeynArts} or on {\it GraphShot}~\cite{GraphShot}
for generating the vertex amplitudes. 
The diagrams containing 
self-energy subloops (e.g.~Fig.~\ref{Higgsdiag}a) 
were evaluated using the method of one-dimensional integral representations, 
as described in~\cite{Hollik:2005va}. 
In addition, as an independent check, 
the methods described in~\cite{Ferroglia:2003yj} 
were implemented and applied. 
The diagrams containing vertex subloops 
(e.g.~Fig.~\ref{Higgsdiag}b) were also
calculated as in the fermionic case, 
using the methods described in~\cite{Ferroglia:2003yj}.
The only new type of diagrams compared to the fermionic case are the two 
non-planar diagrams in Fig.~\ref{Higgsdiag}c,d. The method used for their 
evaluation is explained in the next section. 

\section{Non-planar diagrams}\label{nonplanar}

The non-planar diagrams in Fig.~\ref{Higgsdiag}c,d are UV-finite and can be evaluated 
in 4 dimensions. We have to deal with diagrams of the type
\begin{eqnarray}
V^{\left(\mu,\mu\nu\right)}_{222} &=& 
\int d^4 q_1 \int d^4 q_2 \frac{\left(q_1^\mu,q_1^\mu q_1^\nu\right)}{[1][2][3][4][5][6]},
\end{eqnarray}
with the following short-hand notation for the propagators,
\begin{eqnarray}
&&\nonumber [1]= q_1^2-M^2_V,\qquad[2]= (q_1-p_1)^2,\qquad[3]= (q_1-q_2+p_1)^2-M^2_H,\\
&&\nonumber [4]= (q_1-q_2-p_2)^2-M^2_Z,\qquad[5]= q_2^2,\qquad[6]= (q_2-p_1)^2-M^2_V;\\
&& M_V = \left(M_W,M_Z\right).
\end {eqnarray}
Following the discussion given in \cite{Ferroglia:2003yj}, we first combine the 
propagators [1] and [2] with a Feynman-parameter $z_1$, the propagators [3] and [4] 
with a Feynman-parameter $z_2$ and the propagators [5] and [6] with a 
Feynman-parameter $z_3$. Then we change variables according to 
$q_1 \rightarrow q_1+z_1 \;p_2$, $q_2 \rightarrow q_2+z_3 \;p_1$ 
and combine the $q_1$ and $q_1-q_2$ propagators with a parameter $x$.
After the $q_1$-integration we combine the residual propagators with a 
parameter $y$ and carry out the $q_2$ integration. 

Since we consider the external fermions to be massless we have $p_{1,2}^2=0$. 
In this situation, the expression for $V_{222}$ is much simpler 
than in the general case because we have just to deal with integrals of the type
\begin{eqnarray}
\int^1_0 dx\;dy\;dz_1\;dz_2\;dz_3\;z_3^n\;(a\ z_1\ z_3+b\ z_3+ c\ z_1+d)^{-2}
\end {eqnarray}
with $n = 0,1,2$. 
a,b,c,d are functions of the internal masses, of the external $Z$ momentum
and of the parameters $x,y,z_2$, but they are independent of $z_1$ and $z_3$. 
For $n=0$ we perform the integrations in $z_1$ and $z_3$ analytically,
\begin{eqnarray}
\int^1_0 dz_1\;dz_3\;(a\ z_1\ z_3+b\ z_3+ c\ z_1+d)^{-2} &=& 
\frac{1}{ad-bc} \ln\left(1+\frac{ad-bc}{\left(c+d\right)\left(b+d\right)}\right),
\end {eqnarray}
whereas for $n = 1,2$ the $z_1$ integration and an integration by parts in $z_3$
are done analytically,
\begin{eqnarray}
\nonumber && \int^1_0 dz_1\;dz_3\;z_3^n (a\ z_1\ z_3+b\ z_3+ c\ z_1+d)^{-2} = \\
&&\int^1_0 dz_3 \frac{n z_3^{n-1}}{ad-bc} \ln\left(1+\frac{\left(1-z_3\right)\left(ad-bc\right)}
{\left(\left(a+b\right) z_3+c+d\right)\left(b+d\right)}\right).
\end {eqnarray}
In both cases smooth integrands are obtained, 
suitable for follow-up numerical integrations.
The algebraic handling and the numerical evaluation were performed in two independent 
computations for cross-checking the results.
For the numerical integration
the NAG library D01GDF \cite{naglib} was used  
in one case and the CUBA library \cite{Hahn:2004fe} in the other case.

\section{Results}\label{results}

The evaluation and presentation of the final result are done for the set of
input parameters put together in Tab.~\ref{tab:parameters}. 
$M_W$ and $M_Z$ are the experimental values of 
the $W$- and $Z$-boson masses~\cite{PDG}, 
which are the on-shell masses. 
They have to be converted to the values in the pole mass scheme~\cite{Freitas:2002ja},
labeled as $\overline{M}_W$ and $\overline{M}_Z$, 
which are used internally for the calculation.
These quantities are related via $M_{W,Z} = \overline{M}_{W,Z}+ \Gamma^2_{W,Z}/(2 \;M_{W,Z})$. 
For $\Gamma_Z$ the experimental value (Tab.~\ref{tab:parameters}) and for $\Gamma_W$ the
theoretical value has been used, {\it i.e.}  
$\Gamma_W = 
3 \;G_{\mu} M^3_W/\left(2 \sqrt{2}\pi\right) 
\left(1+ 2 \alpha_s\left(M_W^2\right)/\left(3 \pi\right)\right)$
with sufficient accuracy.

\begin{table}[!htb]
\begin{tabularx}{16.cm}{l c  l}
\hline\hline
parameter & \hspace*{10.cm} &value\\\hline
$M_W$ && $80.426$ GeV\\
$M_Z$ && $91.1876$ GeV\\
$\Gamma_Z$ && $2.4952$ GeV\\
$m_t$ &&  $178.0$ GeV \\
$\Delta\alpha\left(M^2_Z\right)$ && $0.05907$\\
$\alpha_s\left(M^2_Z\right)$ && $0.117$\\
$G_\mu$ &&  $1.16637 \times 10^{-5}$\\
$\overline{M}_W$ && $80.3986$ GeV\\
$\overline{M}_Z$ && $91.1535$ GeV\\\hline\hline
\end{tabularx}
\caption {\small Input parameters entering our computation. $M_W$ and $M_Z$ are the experimental 
values 
of the $W$- and $Z$-boson masses, whereas $\overline{M}_W$ and $\overline{M}_Z$ are 
the calculated 
quantities in the pole mass scheme. }
\label{tab:parameters}
\end{table}

The results are given for $\Delta \kappa$, eq.~(\ref{kappa}), and are listed in 
Tab.~\ref{tab:results} for different values of $M_H$. For comparison, 
Tab.~\ref{tab:results} also contains the values of the fermionic corrections and 
the corresponding subtracted quantity 
$\Delta\kappa^{(\alpha^2)}_{ferm,sub} = \Delta\kappa^{(\alpha^2)}_{ferm}(M_H) -
                                        \Delta\kappa^{(\alpha^2)}_{ferm}(M_H^0)$. 

\begin{table}[!htb]
\begin{tabularx}{16.cm}{c c c c}
\hline\hline
$M_H\left[GeV\right]$& $\Delta\kappa^{(\alpha^2)}_{ferm}\times 10^{-4}$ & $\Delta\kappa^{(\alpha^2)}_{ferm,sub}\times 10^{-4}$ & $\Delta\kappa^{(\alpha^2)}_{bos,sub}\times 10^{-4}$\\\hline
100  &  \hspace*{1.5cm}-0.637(1)  \hspace*{1.5cm} &  \hspace*{1.5cm}0\hspace*{1.5cm}       &  \hspace*{1.5cm}0\hspace*{1.5cm}\\
200  &  \hspace*{1.5cm}-2.165(1)  \hspace*{1.5cm} &  \hspace*{1.5cm}-1.528 \hspace*{1.5cm} &  \hspace*{1.5cm}0.265\hspace*{1.5cm}\\
600  &  \hspace*{1.5cm}-5.012(1)  \hspace*{1.5cm} &  \hspace*{1.5cm}-4.375 \hspace*{1.5cm} &  \hspace*{1.5cm}0.914\hspace*{1.5cm}\\
1000 &  \hspace*{1.5cm}-4.737(1)  \hspace*{1.5cm} &  \hspace*{1.5cm}-4.100 \hspace*{1.5cm} &  \hspace*{1.5cm}1.849\hspace*{1.5cm}\\\hline\hline

\end{tabularx}
\caption {\small Two-loop result for $\Delta \kappa$ in comparison with the fermionic contributions}
\label{tab:results}
\end{table}

In the considered range of the Higgs-boson mass, the bosonic result has the opposite sign and is 
and thus compensates part of the fermionic contributions,
which could be important for the precision expected from the
GigaZ mode of the ILC.
The uncertainties from numerical integration in the bosonic result are of the order $10^{-9}$ 
and hence negligible.

At the end, according to (\ref{kappa}), the final $M_H$-dependence of $\seff$ is obtained in 
combination with $M_W(M_H)$ derived from $G_\mu$ and $\Delta r$ \cite{Awramik:2002wn}. 
The two contributions, from $M_W$ and $\kappa$, have different sign and cancel each other to a large extend, as illustrated in Tab.~\ref{MW-kappa}.

\begin{table}[!htb]
\begin{tabularx}{16.cm}{c c c c}
\hline\hline
$M_H\left[GeV\right]$& $\Delta M_{W,bos}^{(\alpha^2)}\left[MeV\right]$\cite{Awramik:2002wn} & $\seff^{sub}\left(\Delta M_W\right) \times 10^{-5}$ & $\seff^{sub}\left(\Delta \kappa\right)\times 10^{-5}$\\\hline
100  &  \hspace*{1.5cm} -1.0  \hspace*{1.5cm} &  \hspace*{1.5cm}0     \hspace*{1.5cm} &  \hspace*{1.5cm}0   \hspace*{1.5cm}\\
200  &  \hspace*{1.5cm} -0.5  \hspace*{1.5cm} &  \hspace*{1.5cm} -0.97 \hspace*{1.5cm} &  \hspace*{1.5cm}0.59\hspace*{1.5cm}\\
600  &  \hspace*{1.5cm} -0.1  \hspace*{1.5cm} &  \hspace*{1.5cm} -1.74 \hspace*{1.5cm} &  \hspace*{1.5cm}2.03\hspace*{1.5cm}\\
1000 &  \hspace*{1.5cm}  0.6  \hspace*{1.5cm} &  \hspace*{1.5cm} -3.10 \hspace*{1.5cm} &  \hspace*{1.5cm}4.11\hspace*{1.5cm}\\\hline\hline

\end{tabularx}
\caption {\small Variation of  $\seff$ originating 
from $M_W(M_H)$ (column 3) 
in comparison with the variation resulting from $\Delta \kappa$.
Column 2 contains the bosonic 2-loop contributions to $M_W$ from~\cite{Awramik:2002wn}.}
\label{MW-kappa}
\end{table}

In conclusion, we have evaluated the bosonic electroweak 2-loop corrections to $\seff$ 
containing internal Higgs bosons. 
As a new feature, non-planar vertex diagrams appear, and
a method to calculate such non-planar diagrams has been described. 
Our numerical result for $\Delta\kappa^{(\alpha^2)}$ 
shows that the Higgs-mass dependence of the two-loop prediction for $\Delta\kappa$ 
is affected by the bosonic contributions compensating almost
the corresponding contribution to $\seff$  
induced by the bosonic 2-loop corrections to $M_W(M_H)$.
Hence, without the bosonic $\Delta\kappa$ contributions,
the variation of $\seff$ with $M_H$
through the $W$ mass alone would go into the wrong direction.

\vspace*{2cm}
\noindent
S.U. would like to express his gratitude to Stefano Actis for making
available the beta version of GraphShot, a FORM code for generating
and reducing standard model one- and two-loop diagrams which is
presently under development at Torino University.

\end{document}